\documentclass[conference,9pt]{IEEEtran}
\IEEEoverridecommandlockouts
\usepackage{cite}
\usepackage{amsmath,amssymb,amsfonts}
\usepackage{graphicx}
\usepackage{textcomp}
\usepackage{xcolor}
\usepackage{booktabs} 
\usepackage{quantikz}
\usepackage{float}
\usepackage{algorithm,algorithmicx,algpseudocode}
\usepackage{amsmath}  
\usepackage{braket}
\usepackage{multirow}
\usepackage{enumitem}
\usepackage{mathtools}
\usepackage{stfloats}
\usepackage{url}
\usepackage{cleveref}
\usepackage{adjustbox}
\usepackage{array}
\usepackage{caption}
\usepackage{enumitem}

\setlist[itemize]{leftmargin=*}%
\setlist[enumerate]{leftmargin=*}%

\usepackage[compact]{titlesec}

\titlespacing\section{0pt}{0.3\baselineskip}{0.2\baselineskip}
\titlespacing\subsection{0pt}{0.2\baselineskip}{0.1\baselineskip}
\titlespacing\subsubsection{0pt}{0.15\baselineskip}{0.1\baselineskip}

\renewcommand{\baselinestretch}{1.0}

\def\BibTeX{{\rm B\kern-.05em{\sc i\kern-.025em b}\kern-.08em
    T\kern-.1667em\lower.7ex\hbox{E}\kern-.125emX}}

\usepackage{listings}  
\begin{document}

\title{Evaluating Quantum Amplitude Estimation for Pricing Multi-Asset Basket Options}

\author{\IEEEauthorblockN{Muhammad Kashif \IEEEauthorrefmark{1}\IEEEauthorrefmark{2}, Shaf Khalid\IEEEauthorrefmark{1}\IEEEauthorrefmark{2}, Nouhaila Innan\IEEEauthorrefmark{1}\IEEEauthorrefmark{2}, Alberto Marchisio\IEEEauthorrefmark{1}\IEEEauthorrefmark{2},
Muhammad Shafique\IEEEauthorrefmark{1}\IEEEauthorrefmark{2}}

\IEEEauthorblockA{\IEEEauthorrefmark{1}  eBrain Lab, Division of Engineering, New York University Abu Dhabi, PO Box 129188, Abu Dhabi, UAE}
\IEEEauthorblockA{\IEEEauthorrefmark{2} \normalsize Center for Quantum and Topological Systems, NYUAD Research
Institute, New York University Abu Dhabi, UAE}

Emails: \{muhammadkashif, sk10741, nouhaila.innan, alberto.marchisio, muhammad.shafique\}@nyu.edu

\vspace{-10pt}
}

\maketitle

\begin{abstract}
Accurate and efficient pricing of multi-asset basket options poses a significant challenge, especially when dealing with complex real-world data. In this work, we investigate the role of quantum-enhanced uncertainty modeling in financial pricing options on real-world data. Specifically, we use quantum amplitude estimation and analyze the impact of varying the number of uncertainty qubits while keeping the number of assets fixed, as well as the impact of varying the number of assets while keeping the number of uncertainty qubits fixed. To provide a comprehensive evaluation, we establish and validate a hybrid quantum-classical comparison framework, benchmarking quantum approaches against classical Monte Carlo simulations and Black-Scholes methods. Beyond simply computing option prices, we emphasize the trade-off between accuracy and computational resources, offering insights into the potential advantages and limitations of quantum approaches for different problem scales. Our results contribute to understanding the feasibility of quantum methods in finance and guide the optimal allocation of quantum resources in hybrid quantum-classical workflows.
\end{abstract}

\begin{IEEEkeywords}
Pricing Basket Options, Quantum Amplitude Estimation, Multi-Asset Derivatives, Real-World Data
\end{IEEEkeywords}

\section{Introduction}
\label{sec:introduction}

Options are financial derivative contracts that give the buyer the right, but not the obligation, to buy (call option) or sell (put option) an underlying asset at an agreed-upon price (strike) and time frame (exercise window). In their simplest form, the strike price is a fixed value, and the time frame is a single point in
time~\cite{Stamatopoulos_2020}.
%
Basket options are popular derivative contracts that are becoming increasingly widespread in many financial markets, for example, equity, Foreign Exchange, and commodity markets. Given a vector of weights \(\mathbf{w} = (w_1, \ldots, w_n) \in \mathbb{R}^n\), the basket is defined as the weighted arithmetic average of the \(n\) stock prices \(S_1(t), \ldots, S_n(t)\) at time \(T\):
\begin{equation}
    A_n(T) = \sum_{k=1}^n w_k S_k(T).
\end{equation}

Traditional methods, such as classical Monte Carlo simulations, can become computationally expensive for high-dimensional baskets~\cite{Barola_2013}.
Moreover, conventional analytical approaches (e.g., Black–Scholes) often rely on simplifying assumptions (e.g., constant volatility) that fail to capture many real-market complexities~\cite{Jankova_2018}.
While advanced variance-reduction techniques and high-performance computing platforms have improved the runtime of classical methods~\cite{Zhang_2020}, these approaches still face inherent limitations due to their assumptions, such as constant volatility.
Consequently, there is growing interest in exploring \emph{alternate computing paradigms}, such as quantum computing, which operate on a fundamentally different principles and promise significant speedups in certain computational tasks \cite{kashif2025computational}.

Recently, there has been significant progress in exploring quantum-based methods for addressing various finance-related problems~\cite{herman2022survey,innan2024financial23,innan2024qfnn,innan2024financial,dutta2024qadqn,innan2024lep, pathak2024resource,choudhary2025hqnn,alami2024comparative,zaman2024po,innan2025quantum,sawaika2025privacy,el2024quantum}. Quantum amplitude estimation (QAE), which leverages quantum interference and amplitude amplification to estimate probabilities encoded in quantum states, is frequently used for pricing multi-asset basket options, primarily because it promises a quadratic speedup over traditional classical Monte Carlo methods~\cite{Stamatopoulos_2020,Woerner_2019}. 

However, most demonstrations rely on \emph{synthetic} or simplified parameterized models, rather than \emph{fully empirical} market data~\cite{Stamatopoulos_2020}. This leaves open questions regarding QAE's performance on real-world data, particularly when dealing with correlated multi-asset baskets under realistic market conditions. 
In addition, recent works propose resource-optimized approaches for loading distributions~\cite{Alhajjar_2023}, but large-scale demonstrations remain challenging.


\subsection{Related Work}
\subsubsection{Classical Methods and their limitations} 
\paragraph{Black–Scholes Model}
Originally developed for \emph{single-asset} European options \cite{levy2018single}, the Black–Scholes model assumes constant volatility, continuous trading, and lognormal price dynamics. Under these conditions, it provides a closed-form solution for an option’s fair value based on parameters like the current price \(S_0\), strike \(K\), risk-free rate \(r\), and volatility \(\sigma\). Although elegant and computationally efficient for one asset, \emph{multi-asset} extensions of Black–Scholes method have
considerable limitations and constricting
assumptions \cite{Jankova_2018}. 


\begin{figure*}[h]
    \centering
    \includegraphics[width=1.0\linewidth]{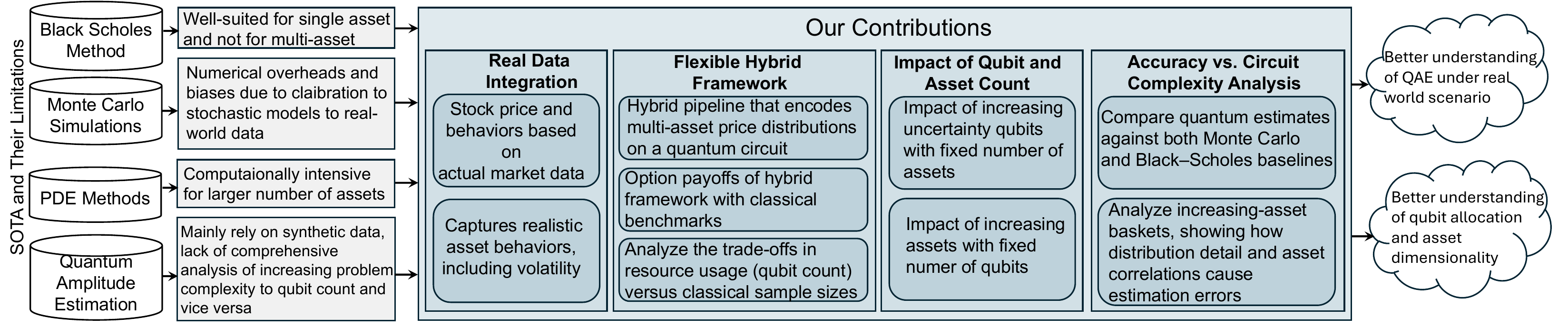}
    \caption{An overview of our contributions.}
    \label{fig:cont}
\end{figure*}

\paragraph{PDE Methods}
A partial differential equation (PDE) approach generalizes the single-asset Black--Scholes PDE to multiple assets, resulting in high-dimensional equations that capture each asset's dynamics and correlations~\cite{Hull_Book,Achdou_Pironneau_2009}
Finite difference or finite element methods can, in principle, be used to numerically solve these PDEs and obtain the option's fair value. However, the computational cost \emph{escalates} quickly with the number of assets. 


\noindent
\paragraph{Monte Carlo Simulations}
In a Monte Carlo (MC) framework, one simulates a large number of possible price paths for each underlying asset (often generated by advanced stochastic processes) and then averages the discounted payoffs to estimate the option’s fair values~\cite{Boyle_1977}. 
MC simulations are popular because they are conceptually simple, relatively easy to implement, and flexible in handling various payoff structures. 
However, they become prohibitively time-consuming for high-dimensional baskets or when seeking tight error bounds, often necessitating variance-reduction techniques or large-scale computational resources~\cite{Barola_2013}.
%
In addition, calibrating stochastic models to real-world market data can introduce biases and numerical overheads~\cite{Andres_2021}.
Hence, these methods often require a large number of samples to capture realistic market features such as volatility clustering, fat-tailed distributions, and cross-asset correlation, leading to high computational costs~\cite{Barola_2013}.

\subsubsection{Quantum Methods and their limitations}

QAE has recently gained attention as a promising approach for improving the efficiency of pricing options~\cite{Stamatopoulos_2020}. 
In principle, QAE enables a \emph{quadratic speedup} over classical Monte Carlo by using the quantum phase estimation paradigm to estimate expectation values (i.e., option payoffs) more rapidly \cite{Stamatopoulos_2020}.
Several works have illustrated how to load probability distributions onto quantum states, approximate payoff functions, and then measure the expected payoff~\cite{Stamatopoulos_2020}. 
Recent research also tackles real-world data embeddings, correlation modeling, and circuit-depth optimizations for multi-asset baskets.

Nevertheless, most demonstrations of QAE rely on either synthetic data or simplified assumptions that overlook real-world market complexities.~\cite{Woerner_2019}. Moreover, existing quantum hardware constraints (limited qubit counts, noisy environments, and circuit-depth limitations) restrict the ability to encode large baskets or refine payoff functions with high accuracy. We also highlight these challenges, showing that while small-scale quantum circuits can approximate basket payoffs, significant discrepancies may arise when transitioning to correlated, higher-dimensional assets (see Section~\ref{sec:results}). Due to limitations in available qubits and circuit depth on current quantum devices, it is essential to systematically analyze how increasing the number of qubits affects performance as the asset portfolio size grows.



\subsection{Our Contributions}
\label{subsec:our_contributions}

In this paper, we propose a comprehensive quantum-classical workflow for pricing multi-asset basket options using \emph{real historical market data}. We then benchmark our quantum results against the classical Monte Carlo and Black–Scholes methods in terms of accuracy. Our key contributions can be summarized as follows:

\begin{itemize}
    \item \textbf{Real-Data Integration:} Unlike state-of-the-art works that rely heavily on synthetic datasets and idealized models, we embed stock prices and behaviors derived from actual market data into quantum states. This approach captures realistic asset behaviors, including volatility and drift, thereby bridging the gap between theoretical QAE demonstrations and practical financial scenarios.

    \item \textbf{Flexible Hybrid Framework:} We establish a \emph{hybrid quantum-classical} pipeline that encodes multi-asset price distributions on a quantum circuit (via uncertainty qubits) and compares the resulting option payoffs with classical benchmarks. This framework highlights the trade-offs in resource usage (qubit count) versus classical sample sizes for different basket sizes and payoff structures.

    \item \textbf{Parametric Studies on Qubit Allocations and Asset Counts:} 
    We systematically vary (1) the number of \emph{uncertainty qubits} per asset while keeping the number of assets fixed, and (2) the number of assets while keeping the uncertainty qubit count fixed. Our analyses reveal the experimental setting where amplitude estimation produces results similar to classical methods, as well as the setting where it deviates and is more accurate than classical benchmarks, providing practical guidance for using current quantum hardware.
    
    \item \textbf{Detailed Accuracy vs. Circuit Complexity Analysis:} 
    We perform extensive comparative analysis of quantum amplitude estimations with classical Monte Carlo and Black–Scholes estimates. One of our key observations is that when we use only a few qubits (e.g., 1–2 qubits), the set of possible asset prices is very basic, which leads to noticeable errors in the final basket price. As we add more uncertainty qubits (e.g., 3–4 qubits), the grid becomes finer, and these errors are reduced. This comes at the cost of exponentially increasing the required computational resources. Specifically, for each additional qubit, the number of grid points doubles, which significantly increases the complexity of the quantum computation. Moreover, when dealing with larger baskets (i.e., more assets), increasing the number of qubits to improve the precision of our calculations can lead to payoff estimates that are either too high or too low. More qubits can help reduce errors, but they also increase the complexity of the calculations.

\end{itemize}

Overall, our work extends state-of-the-art quantum finance studies by combining \emph{real-world data}, \emph{flexible basket modeling}, and a \emph{unified parametric analysis} of qubit allocations and asset dimensionality. This contributes new evidence for the feasibility of quantum approaches in pricing multi-asset basket options, clarifies current hardware limitations, and provides actionable insights for optimizing quantum resources in finance. 


\section{Background}\label{sec:related_work}
Classical pricing basket options has been extensively studied in quantitative finance, with approaches spanning from partial differential equations (PDEs) to Monte Carlo simulations~\cite{Jankova_2018,Barola_2013}. 
PDE-based methods can, in principle, handle multi-asset correlation structures, but they often become computationally expensive or even intractable as the dimensionality (\emph{i.e.}, number of assets) grows large. 
Similarly, Monte Carlo approaches may require millions of simulated paths to capture correlated price behaviors and realistic market features such as stochastic volatility and jumps, leading to high computational overhead~\cite{Barola_2013}. 
Consequently, there is growing interest in exploring alternative paradigms, such as quantum computing, that can potentially reduce runtime while preserving or even improving accuracy.




\subsection{Black–Scholes Approach}
The Black–Scholes (BS) model is a seminal result in quantitative finance for pricing \emph{single-asset} European call options~\cite{Hull_Book}. Under assumptions of constant volatility \(\sigma\), continuous trading, and no market frictions, the BS formula for a call option with current underlying price \(S_0\), strike \(K\), maturity \(T\), and risk-free rate \(r\) is given by\cite{qiskit_basket_option_tutorial}:

\begin{equation}
\label{eq:bs_call}
C = S_0 \, \Phi(d_1) \;-\; K\, e^{-rT}\,\Phi(d_2),
\end{equation}

where \(\Phi(\cdot)\) is the cumulative distribution function (CDF) of the standard normal distribution, and

\begin{align}
d_1 &= \frac{\ln\!\bigl(\tfrac{S_0}{K}\bigr) + \bigl(r + \tfrac{1}{2}\sigma^2\bigr) T}{\sigma \sqrt{T}}, \label{eq:d1} \\
d_2 &= d_1 \;-\; \sigma \sqrt{T}. \label{eq:d2}
\end{align}

Eq.~\eqref{eq:bs_call} originates from the Black–Scholes–Merton partial differential equation, which has an analytical solution under these simplifying assumptions. For a put option, one may use put-call parity or derive a similar expression.
%
Although originally devised for single-asset options, the Black–Scholes model does not easily extend to baskets of multiple assets, especially if asset correlations must be considered. Consequently, practitioners typically resort to numerical methods (e.g., PDEs or Monte Carlo).
%


\subsection{Monte Carlo Simulations}
The classical MC approach simulates a large number of possible paths for each underlying asset (under assumed dynamics, e.g., geometric Brownian motion), calculates the option payoff for each simulated path, and then averages (and discounts) these payoffs to estimate the fair option price:

\begin{equation}
    \label{eq:mc_payoff_estimate}
    \text{Price} \approx e^{-rT} \,\frac{1}{M} \sum_{m=1}^{M} \Bigl[\max\bigl(0,\, A_n^{(m)}(T) - K\bigr)\Bigr],
\end{equation}

where \(M\) is the number of Monte Carlo paths, \(r\) is the risk-free rate, \(A_n^{(m)}(T)\) is the final basket price of the \(m\)-th simulated path at maturity \(T\), and \(K\) is the strike price. While MC is straightforward to implement, it can become computationally very expensive for high-dimensional baskets or for obtaining tight confidence intervals.

\subsection{Quantum Amplitude Estimation}
With the advent of quantum computing, researchers have begun investigating its potential in finance, particularly for pricing options~\cite{Orus_2019}. Building on the concept of \emph{quantum amplitude estimation} (QAE)~\cite{Woerner_2019}, one can in principle achieve a \emph{quadratic} speedup over classical Monte Carlo for estimating expectation values. This speedup relies on using quantum states to represent probability distributions (e.g., of asset prices) and leveraging quantum interference to reduce the number of required samples from \(\mathcal{O}(\varepsilon^{-2})\) to \(\mathcal{O}(\varepsilon^{-1})\) for a target precision \(\varepsilon\)~\cite{Woerner_2019,Stamatopoulos_2020}.

QAE is a quantum algorithm that generalizes \emph{quantum phase estimation} to find the amplitude of a specific outcome in a quantum state~\cite{Stamatopoulos_2020}. Suppose we have a unitary operator \(\mathcal{A}\) that acts on an \emph{initial state} \(\ket{0}^{\otimes n}\) and prepares a state encoding a probability distribution over asset prices:

\begin{equation}
    \label{eq:qae_state_preparation}
    \mathcal{A} \ket{0}^{\otimes n}
      = \sqrt{p}\,\ket{\psi_1}\ket{1} \;+\; \sqrt{1 - p}\,\ket{\psi_0}\ket{0},
\end{equation}

where \(p\) is the “target” probability (or amplitude squared) we wish to estimate, \(\ket{\psi_1}\) and \(\ket{\psi_0}\) are normalized states in the \emph{state qubits}, and the last qubit (often called the \emph{objective qubit}) is marked as \(\ket{1}\) if the payoff condition is satisfied (e.g., basket price above strike).

\paragraph{Grover Operator and Phase Estimation}
To estimate \(p\), QAE applies a form of amplitude amplification (via a Grover-like operator) multiple times and uses an inverse Quantum Fourier Transform (QFT) or iterative methods to extract the phase that encodes \(p\). In standard QAE:
\begin{equation}
   p = \sin^2 \bigl(\theta\bigr), \quad \text{where} \quad \theta \approx \frac{\varphi}{2^d},
\end{equation}
and \(\varphi\) is obtained from phase estimation with \(d\) ancilla qubits. Alternatively, Iterative Amplitude Estimation (IAE)~\cite{Woerner_2019} avoids the full QFT by incrementally refining the estimate of \(p\).

\paragraph{Uncertainty and State Qubits}
In pricing basket options, each asset’s future price distribution is discretized into a set of bins, and these bins are loaded into a register of \emph{uncertainty qubits}~\cite{Woerner_2019,Stamatopoulos_2020}. For instance, if one allocates \(n\) qubits to each asset, the dimension of the probability space is \(2^n\) per asset. The combined distribution for a basket of \(d\) assets could then require \(d \times n\) qubits to represent. An additional \emph{objective qubit} is typically used to mark whether or not the payoff condition (e.g., \(\max(0,\text{basketPrice} - K)\)) is satisfied, and then the seller can sell or the buyer can buy.




\begin{figure*}
    \centering
    \includegraphics[width=1\linewidth]{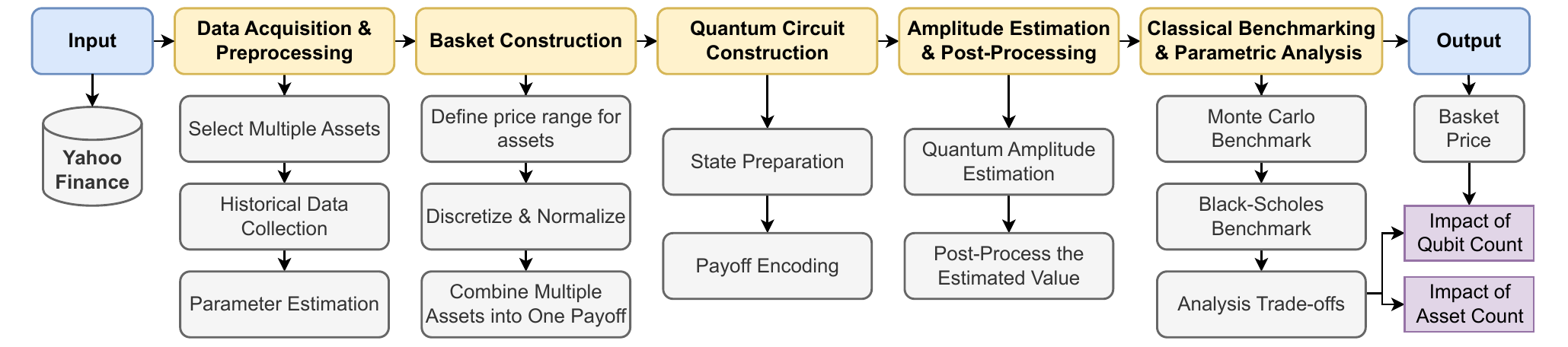}
    \caption{Our methodology for pricing multi-asset basket options using Quantum Amplitude Estimation.}
    \label{fig:method_flow}
\end{figure*}


\section{Our Methodology}
\label{sec:methodology}

In this paper, we consider the problem of pricing multi-asset basket options using QAE. Our methodology integrates real-world financial data with quantum state preparation techniques to model the probability distribution of asset prices. We construct a quantum circuit that first loads the distributions (via uncertainty qubits) for each asset. Afterwards, by varying the number of uncertainty qubits and the basket dimensionality, we analyze the trade-offs for estimation accuracy. Our approach is benchmarked against classical Monte Carlo simulations and Black–Scholes approximations to assess its practical viability.
Fig.~\ref{fig:method_flow} summarizes our pipeline for pricing multi-asset basket options using QAE. 
%
%

\subsection{Data Acquisition \& Preprocessing}
\label{subsec:method_stage1}

The first step in our methodology involves selecting a set of \(d\) assets (e.g., AAPL, GOOG, MSFT) along with their corresponding weights \((w_1, w_2, \ldots, w_d)\), which represent the contribution of each asset to the basket option. We retrieve the daily \textit{Adjusted Close} prices for each asset over a specified time interval \([t_0, t_1]\) using the \texttt{yfinance} Python library. This library provides easy access to historical stock data and handles missing values, such as those due to non-trading days (e.g., weekends, market holidays). Any missing data points are excluded from the dataset to ensure the integrity of the time series. We consider real historical data from January $1^{st}$ to June $30^{th}$ $2024$.

For each asset \(i\), let \(S_{i,t}\) represent the price of asset \(i\) on day \(t\). We then compute the \textit{logarithmic returns}, \(r_{i,t}\), for each asset over consecutive days using the formula:
\[
r_{i,t} = \ln\left(\frac{S_{i,t}}{S_{i,t-1}}\right),
\]
where \(S_{i,t-1}\) is the price of asset \(i\) on the previous day. Logarithmic returns are often used in financial modeling because they are time additive and provide a more straightforward way to model returns over multiple periods.

Next, we estimate each asset’s \textit{mean} return \(\mu_i\) and \textit{standard deviation} \(\sigma_i\) based on the historical returns. These estimates are then scaled to account for the desired \textit{maturity} \(T\) of the option, which is expressed in years. The scaled values for mean and standard deviation are computed as:
\[
\mu_{i,T} = \mu_i \cdot T, \quad \sigma_{i,T} = \sigma_i \cdot \sqrt{T}.
\]
We add a small shift to the mean return \(\mu_i\) (e.g., \(+5\%\)) to reflect market optimism. 
Once the estimates for the mean and standard deviation are computed, we then compute the initial \textit{basket price} at time \(t = 0\) using the weighted sum of the adjusted close prices:
\[
\text{BasketPrice}_0 = \sum_{i=1}^{d} w_i \, S_{i,0},
\]
where \(S_{i,0}\) is the adjusted close price of asset \(i\) on the first day of the dataset, and \(w_i\) is the weight of asset \(i\) in the basket. This initial price serves as the starting point for calculating future basket prices and their associated payoffs.
Finally, the adjusted prices, log returns, and the initial basket price are stored and used for further analysis.

\subsection{Distribution Discretization \& Basket Construction}
\label{subsec:method_stage2}

\paragraph{Individual Asset Grids}
For each asset \(i\), we define the price range as in \Cref{eq:price_range}.

\begin{equation}
\begin{cases}
S_{\min}^{(i)} = S_{i,0} \exp\left(\mu_{i,T} - 3\,\sigma_{i,T}\right) \\S_{\max}^{(i)} = S_{i,0} \exp\left(\mu_{i,T} + 3\,\sigma_{i,T}\right)
\end{cases}
\label{eq:price_range}
\end{equation}

This range is divided into \(2^{n_i}\) points, where \(n_i\) represents the number of uncertainty qubits for asset \(i\). We then evaluate the lognormal probability density function (PDF) at these points and normalize it to get probabilities \(\{p_{i,0}, p_{i,1}, \ldots\}\).

\paragraph{Combining into a Basket Price Grid}
Assuming the assets are independent,\footnote{Our current implementation focuses on independent assets; correlation modeling may be incorporated with more complex state-preparation strategies.} we create a multi-dimensional grid for all \(d\) assets. Each grid point corresponds to a vector \((S_1, S_2, \ldots, S_d)\), where the probability of each vector is the product of individual asset probabilities. We then calculate the basket price at each grid point:
\[
A_d(T) = \sum_{i=1}^{d} w_i S_i,
\]
where \(w_i\) is the weight of asset \(i\). These basket prices are then binned into a 1D array of size \(2^{\textit{nbq}}\), where \(\textit{nbq}\) is the number of qubits allocated to represent the basket price. The result is a discretized basket price distribution with probabilities \(\{\pi_0, \pi_1, \ldots, \pi_{2^\textit{nbq}-1}\}\).

\paragraph{Price Grid and Discretization}
A price grid is a representation of possible prices for an asset or a basket of assets over a given range. It helps in analyzing the changes in prices over time or under different conditions. To analyze these possible prices, we break the range into several discrete points, which form the grid.

Discretization refers to the process of dividing a continuous range of values into a finite set of intervals. For example, if we have a range of stock prices between \$100 and \$200, we can discretize that range into a set of smaller points like \$100, \$110, \$120, and so on. This makes the range manageable, as we can work with a limited number of values instead of an infinite number.

Discretization is a main step when working with quantum algorithms as they require a finite set of discrete values to process and calculate probabilities. Quantum systems can only represent and manipulate discrete states, so continuous values like asset prices must be converted into a manageable number of discrete intervals. This allows quantum circuits to perform operations on these values efficiently, making it possible to compute outcomes while ensuring the system remains within computational limits. 

\subsection{Quantum Circuit Construction} 
\label{subsec:method_stage3}

After preparing the financial data and discretizing the asset prices, we construct a quantum circuit to encode the probability distribution of the basket price. In our implementation, the circuit is designed for a basket option with two underlying assets whose prices at maturity are assumed to follow independent log-normal distributions. The construction involves two main steps: \emph{state preparation} of the underlying uncertainty model and \emph{payoff encoding} via reversible arithmetic and controlled rotations.

\paragraph{State Preparation}

To encode the uncertainty of the basket price, we begin with the classical probability distribution for the asset prices. For each asset, the continuous log-normal distribution is first truncated to an interval \([l, h]\) and then discretized into \(2^n\) grid points using \(n\) uncertainty qubits. The affine mapping that converts a discretized integer \(x\) to the continuous interval is given by:
\begin{equation}
x \mapsto l + \frac{x}{2^n-1} (h-l).
\end{equation}
In our implementation, we use the \texttt{LogNormalDistribution} to generate the quantum state that represents this discretized distribution; the function is instantiated as
\begin{equation}
\begin{aligned}
u =\ & \text{LogNormalDistribution}(num\_qubits, mu, sigma=cov, \\
     & \quad bounds=list(zip(low, high)))
\end{aligned}
\end{equation}
where \(\texttt{num\_qubits}\) is a list containing the number of uncertainty qubits per asset; $\mu$ is the mean (in log-space) and \(\sigma\), which represents covariance matrix \(\sigma^2 I\) for independence, is the volatility-related parameter; and \(\texttt{bounds}\) is defined by the interval \([l, h]\) for each asset.

The resulting quantum state is stored in the \texttt{qr\_state} register, and its amplitudes correspond to the square roots of the discretized probabilities,
\begin{equation}
\alpha_j = \sqrt{\pi_j}, \quad \text{with} \quad \sum_j \alpha_j^2 = 1.
\end{equation}
This prepares the state such that measuring \(\texttt{qr\_state}\) yields the basket price with probability \(\pi_j\).
\paragraph{Payoff Encoding}
The basket option payoff is defined classically as:
\begin{equation}
\text{Payoff} = \max \bigl(0, A_d(T) - K \bigr),
\end{equation}
where \(A_d(T)=S_T^1+S_T^2\) is the aggregated asset price at maturity, and \(K\) is the strike price. In our quantum circuit, after state preparation, we need to map the price information into a single \emph{objective qubit} that encodes the (scaled) payoff.

Since the weighted sum operator available in our framework only handles integer inputs, we first apply a \emph{weighted adder} using the \texttt{WeightedAdder}\footnote{\url{https://docs.quantum.ibm.com/api/qiskit/qiskit.circuit.library.WeightedAdder}}. The weights are chosen according to the binary expansion:
\begin{equation}
w_{i} = 2^{i} \quad \text{for } i=0,\ldots,n-1,
\end{equation}
and the adder circuit computes the integer sum of the discretized asset prices. Let \(n_s\) denote the number of qubits required to represent the sum.

The strike price \(K\) is then mapped from the interval \([l, h]\) to the integer domain \(\{0,\dots,2^n-1\}\) via
\begin{equation}
K_{\text{mapped}} = \left( \frac{K - d\,l}{h - l}\right) \Bigl(2^{n} - 1\Bigr),
\end{equation}
where \(d\) is the number of assets. This transformation aligns the strike price with the discretized output of the weighted adder.

A piecewise-linear function is employed to approximate the payoff function. In practice, this is implemented via \texttt{LinearAmplitudeFunction}, which models the mapping
\begin{equation}
f(x) = \begin{cases}
0,& x < K_{\text{mapped}}, \\
c\, (x - K_{\text{mapped}}),& x \ge K_{\text{mapped}},
\end{cases}
\end{equation}
where \(c\) is an approximation scaling factor. This function is embedded into the circuit via controlled rotations; that is, after computing the sum (stored in an ancilla register, denoted \(\ket{\text{sum}}\)), the operator applies a rotation to the \texttt{qr\_obj} (objective qubit) such that its amplitude corresponds to the scaled payoff.



\subsection{Amplitude Estimation \& Post-Processing}
\label{subsec:method_stage4}

In this step, we combine the \emph{state-preparation} and \emph{payoff} circuits into an \texttt{EstimationProblem}. The goal is to refine the probability of the objective qubit being in the ``payoff'' state with a target precision \(\epsilon_\text{target}\) and confidence \(\alpha\). QAE is used to estimate the expected payoff:
\begin{equation}   
\mathbb{E}\Bigl[\max\bigl(0,\,A_d(T) - K\bigr)\Bigr],
\end{equation}
where \(A_d(T)\) is the basket price at maturity and \(K\) is the strike price. This expected payoff is calculated on a discretized grid and then scaled by the chosen factor.

After obtaining the result, we map it back to the original basket-price range and subtract \(K\) if needed or just keep the final basket price, to account for the strike price. The precision and confidence of this estimation depend on the chosen parameters, \(\epsilon_\text{target}\) and \(\alpha\), respectively.


\subsection{Classical Benchmarking \& Parametric Analysis}
\label{subsec:method_stage5}

In this section, we compare the quantum-based estimates to several classical reference methods to evaluate their performance:

\begin{itemize}[leftmargin=*]
    \item \textbf{Classical Binned Summation (Final Basket Price):} This method involves calculating the payoff for each possible price point in the discretized distribution (the same distribution used for quantum encoding) and summing them to get the total payoff.
    \item \textbf{Monte Carlo (MC) Simulation:} This technique generates a large number \(M\) (e.g., 10,000) of lognormal samples for each asset, computes the payoff \(\max(0, A_d(T) - K)\) for each sample, and averages the results to estimate the expected payoff.
    \item \textbf{Black-Scholes (BS):} For simpler cases or lower-dimensional baskets, we may use the Black-Scholes model to calculate the option price, under certain simplifying assumptions.
\end{itemize}

We also perform \emph{parametric analysis} to investigate how different choices impact the results:

\begin{itemize}
    \item \textbf{Number of Uncertainty Qubits (\(n\)):} Increasing \(n\) leads to finer resolution in the price grid but results in deeper quantum circuits. This trade-off is essential when considering computational resources.
    \item \textbf{Time Horizon (\(T\)):} Changing the time horizon alters the asset’s mean return \(\mu_{i,T}\) and volatility \(\sigma_{i,T}\), affecting the accuracy of the price estimates.
\end{itemize}



\subsection{Final Results and Evaluation}
\label{subsec:method_stage6}

In this section, we evaluate the effectiveness and accuracy of our quantum-based pricing model for multi-asset basket options, with a particular focus on the trade-offs related to uncertainty qubits and asset dimensionality. This evaluation is essential for understanding how well the quantum results align with classical benchmarks and how various parameters affect both accuracy and computational complexity.

\paragraph{Impact of Uncertainty Qubits}

The first factor we evaluate is the number of uncertainty qubits allocated to each asset. As the number of qubits increases, the discretization of the price grid improves, allowing for more accurate QAE. However, this improvement comes at the cost of \textbf{increased quantum circuit depth}, which makes the problem more computationally expensive. We assess the results by comparing the quantum estimates at different qubit levels to the MC simulations and BS model estimates, which serve as classical benchmarks.

In particular, we analyze how the convergence of quantum estimates behaves as the qubit count increases from 1 to 4. For smaller baskets (e.g., 3–4 assets), we expect significant improvements in accuracy with 3–4 qubits, aligning closely with classical methods. However, for larger baskets (e.g., 6–9 assets), we anticipate that increasing the number of qubits further will yield only marginal improvements, as the \textit{“flattening” effect} sets in. This effect highlights the point at which increasing qubits no longer contributes to substantial improvements in pricing accuracy.

\paragraph{Impact of Asset Dimensionality}

The second factor we evaluate is the number of assets in the basket. As the number of assets increases, the \textbf{complexity} of the problem grows, which requires additional quantum resources and increases the difficulty of capturing asset correlations. In this section, we investigate how the quantum model handles different basket sizes, from small baskets with 3 assets to larger baskets with up to 9 assets.

The results are expected to show that for smaller baskets, the quantum model performs similarly to classical Monte Carlo and Black-Scholes estimates, especially with 3–4 qubits. However, as the basket size increases, the quantum estimates are likely to exhibit more \textbf{overshooting} due to limitations in capturing correlations between assets under the assumption of asset independence. We evaluate the trade-offs between discretization fidelity and quantum resource limitations, and discuss potential improvements to address these challenges.

\vspace{5pt}

\noindent



\section{Results and Discussion}\label{sec:results}

We now present the results of our analysis individually on the impact of increasing the number qubits for fixed assets and increasing the assets for fixed qubit count \footnote{From now on, qubits and uncertainty qubits will be used interchangeably}. It is important to note here that we consider the \emph{final basket price} at maturity rather than the option payoff. This is because once the final basket price is known, the payoff can simply be calculated using the following equation:
\begin{equation}
\max \bigl(0,\, \text{(final basket price)} - K\bigr),
\end{equation}
where \(K\) is the strike price. Thus, analyzing the final basket price is sufficient to compare the performance of QAE against classical benchmarks.


\subsection{Impact of Increasing Qubits in QAE and Comparison with Classical Benchmarks}

We start by analyzing the effect of increasing the number of qubits for a fixed number of assets. For each fixed asset configuration, we begin with the minimum number of qubits required for the underlying problem and progressively increase the qubit count until additional qubits yield no significant improvement in accuracy or lead to a degradation in performance.




\begin{figure*}
    \centering
    \includegraphics[width=1\linewidth]{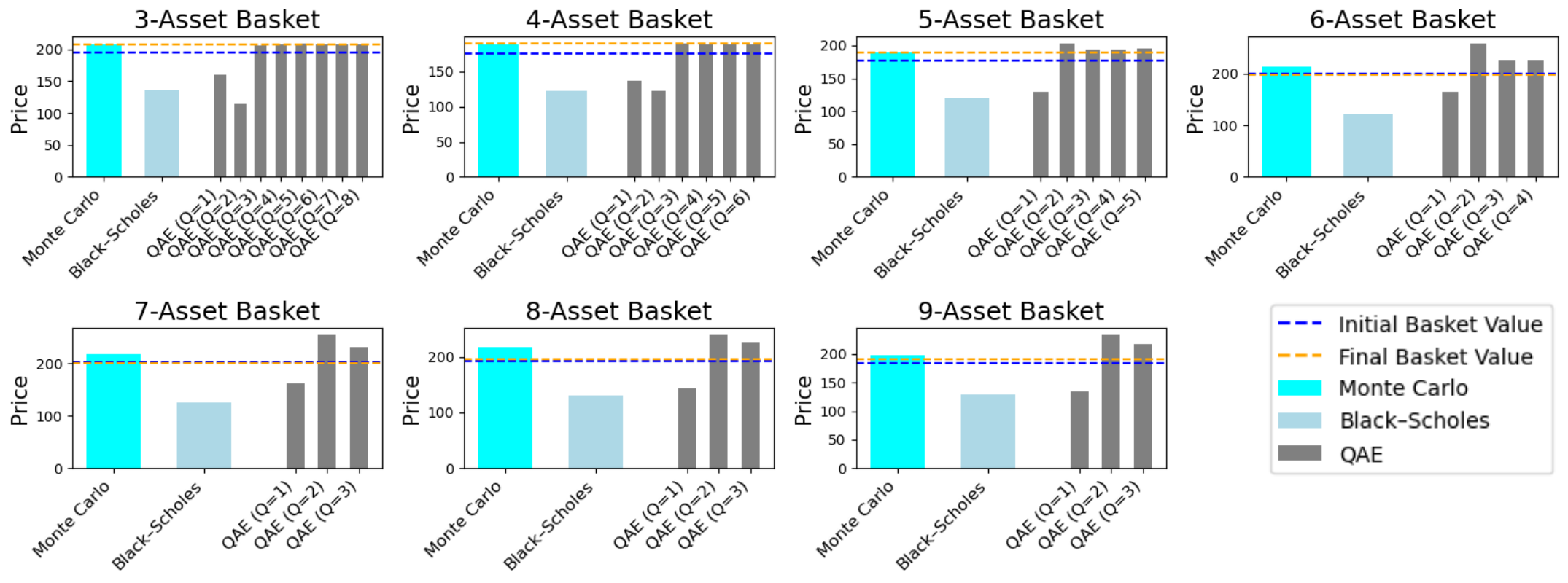}
    \caption{Impact of increasing the qubits while keeping the assets fixed.}
    \label{fig:results_qubit_impact}
\end{figure*}


\subsubsection{Three-Asset Basket}
\label{subsec:results_3_assets}

For the basket of three assets, we consider AAPL, GOOG, MSFT stocks with weights $0.4, 0.4$, and $0.2$ respectively. The initial Basket Price is $\$194.50$ and final basket price is $\$207.16$.
%
%
The results are shown in Fig. \ref{fig:results_qubit_impact}. We observe that when only $1$ or $2$ uncertainty qubits are allocated, the discretization of the basket price distribution is extremely coarse, yielding significant underestimates (with QAE values of $\$160.62$ and $\$113.55$, respectively). With $3$ uncertainty qubits the QAE estimate improves markedly to $\$205.77$, approaching the values of the Monte Carlo expected final basket price ($\$208.25$). As the number of uncertainty qubits increases from $4$ to $8$, the QAE estimates converge steadily, with the best estimate achieved at $8$ qubits being $\$208.2440$. Thus, at its best the QAE method is essentially on par with the Monte Carlo and better than the Black--Scholes approaches, providing a reliable estimate of the final basket price.

\subsubsection{Four-Asset Basket}
\label{subsec:results_4_assets}

For the basket of four assets, we consider AAPL, GOOG, MSFT and NVDA stocks with weights $0.35, 0.30, 0.20, 0.15$. The initial Basket Price is $\$175.61$ and final basket price is $\$190.99$.
%
%
%
The results are shown in Fig. \ref{fig:results_qubit_impact}. We notice that When only $1$ or $2$ uncertainty qubits are allocated, the discretization of the basket price distribution is extremely coarse, leading to significant underestimation (with QAE values of $\$137.18$ and $\$122.69$, respectively). With $3$ uncertainty qubits, the QAE estimate improves markedly to $\$189.56$, which is very close to both the Monte Carlo expected final basket price ($\$187.97$) and better than the Black--Scholes final basket price ($\$123.17$). As the number of uncertainty qubits increases from $4$ to $6$, the QAE estimates converge steadily, with the best estimate at $6$ qubits being $\$188.38$. Thus, with a sufficient number of uncertainty qubits, the QAE method provides an estimate that is effectively on par (but slightly better) than the Monte Carlo approach.


\subsubsection{Five-Asset Basket}
\label{subsec:results_5_assets}
For the basket of five assets, we consider AAAPL, GOOG, MSFT, NVDA and AMZN stocks with weights $0.30, 0.25, 0.20, 0.15, 0.10$. The initial Basket Price is $\$177.03$ and final basket price is $\$189.6$.
%
%
%
The results are presented in Fig. \ref{fig:results_qubit_impact}. We observe that the Monte Carlo simulation estimates the expected final basket price at approximately $\$189.55$, and the Black-Scholes method gives a final basket price of about $\$120.68$. When only $1$ uncertainty qubit is allocated, the QAE method severely underestimates the basket price (yielding $\$129.50$). With $2$ uncertainty qubits, the QAE result overshoots at $\$203.03$. However, as the number of uncertainty qubits increases to $3$, the QAE estimate improves to $\$194.12$ and then converges to around $\$194.50$ when using $4$ or $5$ qubits. Although these converged QAE estimates are higher than both the final basket value and the Monte Carlo expected price, the trend shows that increasing the number of uncertainty qubits refines the discretization, leading to more stable estimates.

\subsubsection{Six-Asset Basket}
\label{subsec:results_6_assets}
For the basket of six assets, we consider AAPL, GOOG, MSFT, NVDA, AMZN and TSLA stocks with weights $0.25, 0.20, 0.20, 0.15, 0.10, 0.10$. The initial Basket Price is $\$199.56$ and final basket price is $\$197.87$.
%
%
%
The results showing the effect of increasing qubits in QAE together with comparative analysis with classical benchmarks are presented in Fig. \ref{fig:results_qubit_impact}. The Monte Carlo simulation estimates the expected final basket price at approximately $\$213.43$, while the Black-Scholes method yields a final basket price of about $\$122.19$. With only $1$ uncertainty qubit, the QAE estimate is significantly low at $\$164.44$. Allocating 2 uncertainty qubits causes the QAE estimate to overshoot drastically to $\$258.03$. Increasing the number of uncertainty qubits to $3$ improves the QAE estimate to $\$224.50$, and with 4 qubits it marginally increases to $\$224.70$. Thus, although the QAE estimates tend to converge with more uncertainty qubits, in this case the converged estimates are higher than both the Monte Carlo and Black-Scholes results, highlighting the sensitivity of the QAE method to the discretization resolution in a six-asset basket setting.

\subsubsection{Seven-Asset Basket}
\label{subsec:results_7_assets}
For the basket of seven assets, we consider AAPL, GOOG, MSFT, NVDA, AMZN, TSLA, and V stocks with weights $0.22, 0.18, 0.18, 0.15, 0.10, 0.10, 0.07$. For the seven-asset basket, the initial basket price is $\$203.01$ and the final basket value is $\$199.93$.
%
%
The results showing the effect of increasing qubits in QAE together with comparative analysis with classical benchmarks are presented in Fig. \ref{fig:results_qubit_impact}. We observe that the Monte Carlo simulation estimates the expected final basket price at approximately $\$217.32$, while the Black-Scholes method yields a final basket price of about $\$126.24$. With only $1$ uncertainty qubit, the QAE estimate is significantly low at $\$161.78$. Allocating $2$ uncertainty qubits results in a substantial overshoot to $\$254.58$. Increasing the number of uncertainty qubits to 3 improves the QAE estimate to \$231.86. Thus, although the QAE estimates tend to converge with an increasing number of uncertainty qubits, in this case the converged values remain higher than both the Monte Carlo and Black-Scholes results, underscoring the sensitivity of the QAE approach to the discretization resolution in a seven-asset basket setting.

\subsubsection{Eight-Asset Basket}
\label{subsec:results_8_assets}
For the basket of eight assets, we consider AAPL, GOOG, MSFT, NVDA, AMZN, TSLA, V, and JNJ stocks with weights $0.20, 0.16, 0.16, 0.14, 0.10, 0.08, 0.08, 0.08$. The initial basket price is $\$192.23$ and the final basket value is $\$195.02$. 
%
%
The results showing the effect of increasing qubits in QAE together with comparative analysis with classical benchmarks are presented in Fig. \ref{fig:results_qubit_impact}. The Monte Carlo simulation estimates the expected final basket price at approximately $\$217.32$, while the Black-Scholes method yields a final basket price of about $\$129.89$. With only $1$ uncertainty qubit, the QAE estimate is significantly low at $\$142.31$. Allocating $2$ uncertainty qubits causes the QAE estimate to overshoot dramatically to $\$238.97$, and with $3$ uncertainty qubits it converges to $\$225.87$. Thus, although the QAE method shows signs of convergence with increasing uncertainty qubits, the final QAE estimate remains higher than both the Monte Carlo and Black-Scholes results, emphasizing the method's sensitivity to discretization resolution in an $8$-asset basket setting.

\subsubsection{Nine-Asset Basket}
For the basket of nine assets, we consider AAPL, GOOG, MSFT, NVDA, AMZN, TSLA, V, JNJ, and XOM stocks with weights $0.18, 0.15, 0.15, 0.12, 0.10, 0.08, 0.08, 0.07, 0.07$. The initial basket price is $\$184.21$ and the final basket value is $\$190.27$.

%
%
The results showing the effect of increasing qubits in QAE together with comparative analysis with classical benchmarks are presented in Fig. \ref{fig:results_qubit_impact}. We notice that the Monte Carlo simulation estimates the expected final basket price at approximately \$197.12, while the Black-Scholes method yields a final basket price of about \$129.29 . With only 1 uncertainty qubit, the QAE estimate is significantly low at \$135.10. Allocating 2 uncertainty qubits overshoots the estimate to \$233.05, and with 3 uncertainty qubits the QAE result converges to \$217.26. Thus, although the QAE estimates show a trend toward convergence as the number of uncertainty qubits increases, the converged value remains higher than both the Monte Carlo and Black-Scholes results, underscoring the sensitivity of the QAE method to the discretization resolution in a 9-asset basket setting.


In all basket sizes (3--9 assets), QAE with only 1--2 qubits per asset suffers from severe under- or overshooting due to coarse discretization. However, at \textbf{3--4 qubits per asset}, QAE’s final basket price in most cases aligns more closely with the final basket price than either Monte Carlo or Black--Scholes. However, beyond 4 qubits, the improvements often taper off, highlighting the trade-off between the improvements we get in accuracy when we increase the number of qubits and the rapidly growing circuit complexity on near-term quantum hardware.

Overall, these experiments demonstrate that, with a sufficient number of uncertainty qubits, the amplitude estimation method reliably converges similar to that of classical solutions, and in some cases even aligns more closely with a particular simulated outcome. When pricing larger baskets (6--9 assets), the complexity increases, demanding a higher number of uncertainty qubits. Although allocating 3--4 uncertainty qubits per asset generally yields reasonable estimates, the final QAE results can still overshoot or undershoot the classical benchmarks. Improved state-preparation techniques and more accurate correlation modeling may further reduce these discrepancies.

In summary, our findings indicate that an allocation of \emph{3--4 uncertainty qubits per asset} represents an attractive near-term sweet spot, effectively balancing circuit depth against the accuracy of pricing multi-asset basket options.

\begin{table*}[h]
    \centering
    \caption{Percentage deviation between actual final value of the basket and estimated value by QAE for varying number of uncertainty qubits.}
    \begin{tabular}{|c|c|c||c|c|c||c|c|c|}
        \hline
        \multicolumn{3}{|c||}{\textbf{1-qubit}} & \multicolumn{3}{c||}{\textbf{2-qubits}} & \multicolumn{3}{c|}{\textbf{3-qubits}} \\
        \hline
        Assets & Final Basket Price & \% Deviation & Assets & Final Basket Price & \% Deviation & Assets & Final Basket Price & \% Deviation \\
        \hline
        3 & 207.16 & 22.70 & 3 & 207.16 & 45.19 & 3 & 207.16 & 0.67 \\
        4 & 190.99 & 28.17 & 4 & 190.99 & 35.76 & 4 & 190.99 & 0.75 \\
        5 & 189.60 & 31.70 & 5 & 189.60 & 7.08 & 5 & 189.60 & 2.38 \\
        6 & 197.87 & 16.90 & 6 & 197.87 & 30.40 & 6 & 197.87 & 13.46 \\
        7 & 199.93 & 19.08 & 7 & 199.93 & 27.33 & 7 & 199.93 & 15.97 \\
        8 & 195.02 & 27.03 & 8 & 195.02 & 22.54 & 8 & 195.02 & 15.82 \\
        9 & 190.27 & 28.99 & 9 & 190.27 & 22.49 & 9 & 190.27 & 14.19 \\
        \hline
    \end{tabular}
    \vspace{10pt}
    \begin{tabular}{|c|c|c||c|c|c|}
        \hline
        \multicolumn{3}{|c||}{\textbf{4-qubits}} & \multicolumn{3}{c|}{\textbf{5-qubits}} \\
        \hline
        Assets & Final Basket Price & \% Deviation & Assets & Final Basket Price & \% Deviation \\
        \hline
        3 & 207.16 & 0.11 & 3 & 207.16 & 0.11 \\
        4 & 190.99 & 1.41 & 4 & 190.99 & 1.41 \\
        5 & 189.60 & 2.58 & 5 & 189.60 & 2.58 \\
        6 & 197.87 & 13.56 & 6 & 197.87 & 13.56 \\
        \hline
    \end{tabular}
    \label{tab:qae_combined_deviation}
\end{table*}

\subsection{Impact of Increasing Asset Dimensionality }
We now analyze the effect of increasing the number of assets while keeping the number of uncertainty qubits fixed. 
Table~\ref{tab:qae_combined_deviation} presents the deviation percentage between the actual and QAE-predicted basket values across varying asset sizes and fixed qubit counts. Analyzing the results, it becomes evident that, for a fixed number of qubits, the accuracy generally deteriorates as the number of assets increases. This trend highlights a fundamental limitation arising from attempting to represent increasingly complex financial states with a fixed quantum resource budget.

For instance, with a single uncertainty qubit, the percentage deviation significantly worsens from 22.70\% for 3 assets to 31.70\% for 5 assets. This deterioration is indicative of insufficient quantum capacity to accurately encode and estimate states associated with larger asset numbers. Conversely, as we allocate additional uncertainty qubits, the deviation reduces substantially, reflecting improved representational capability. Notably, increasing the uncertainty qubit count to 3 and above leads to substantial accuracy improvements, especially at smaller asset counts (e.g., deviations of less than 1\% for 3 and 4 assets).

However, even with higher uncertainty qubit counts, accuracy gains begin to plateau as the asset size continues to grow, demonstrating diminishing returns from adding qubits beyond a certain threshold, particularly for 4 and 5 qubits scenarios. This suggests that optimal resource allocation—balancing the number of qubits against the complexity introduced by the number of assets—is crucial to maximize the efficiency and accuracy of quantum amplitude estimation methods in financial applications.

\section{Conclusion}
\label{sec:conclusion}

We presented an extended, hybrid quantum-classical approach for pricing multi-asset basket options using real-world financial data. By incorporating correlated lognormal models into QAE, we evaluated baskets ranging from three to nine assets and analyzed how the number of uncertainty qubits per asset affects accuracy. Our results show that at low Qubit Counts (1--2) lead to coarse discretizations, causing significant under- or overestimations compared to classical Monte Carlo or Black–Scholes references. However, moderate qubit counts (3--4) bring QAE estimates much closer to classical benchmarks across all tested basket sizes, indicating a sweet spot for near-term quantum hardware. Furthermore, higher qubit counts (beyond 4) result in diminishing returns in accuracy, while sharply increasing circuit depth and resource demands.

Overall, our findings demonstrate that amplitude estimation can reliably converge toward classical valuations when the asset-price grid is sufficiently resolved. However, the persistent discrepancies in larger baskets underscore the need for refined state preparation, and enhanced correlation modeling. We anticipate that future advances in quantum hardware, along with innovative circuit design and robust error mitigation strategies, will further narrow the gap between quantum-based and classical pricing options in high-dimensional financial applications.

\section*{Acknowledgment}
This work was supported in part by the NYUAD Center for Quantum and Topological Systems (CQTS), funded by Tamkeen under the NYUAD Research Institute grant CG008.

\bibliographystyle{IEEEtran}
\bibliography{IEEE}

\end{document}